\documentstyle[epsfig,psfig]{mn2e}

\title[H-ATLAS: A Candidate High Redshift Cluster/Protocluster of Star-Forming Galaxies]{H-ATLAS: A Candidate High Redshift Cluster/Protocluster of Star-Forming Galaxies}
\author[D.L. Clements et al.]
        {\small D.L. Clements$^1$, F. Braglia$^1$, G. Petitpas$^{19}$, J. Greenslade,$^1$, A. Cooray$^9$, E. Valiante,$^2$,  G. De Zotti$^{3,4}$,  B. O'Halloran$^1$, J. Holdship$^1$, B. Morris$^1$, \newauthor
\small        I.~P{\'e}rez-Fournon$^{20,21}$, D. Herranz$^5$, D. Riechers$^{6}$, M. Baes$^{7}$, M. Bremer$^8$, N. Bourne$^{14}$, H. Dannerbauer$^{10}$,  A. Dariush$^{1, 11}$, L. Dunne$^{25,12}$, \newauthor
\small            S. Eales$^2$, J. Fritz$^{7, 29}$, J. Gonzalez-Nuevo$^{5,30}$, R. Hopwood$^{1, 13}$, E. Ibar$^{24}$, R.J. Ivison$^{26, 12}$, L.L. Leeuw$^{16}$, S. Maddox$^{25,12}$, \newauthor
\small             M.J. Micha{\l}owski$^{7, 12}$ M. Negrello$^{3}$,  A. Omont$^{17}$, I.Oteo$^{12,26}$, S. Serjeant$^{13}$, I. Valtchanov$^{28}$, J.D. Vieira$^{22,23}$, J. Wardlow$^{9, 27}$,  P. van der Werf$^{18}$\vspace{0.4cm}\\
             \parbox{\textwidth}{\raggedright
{\tiny	 $^1$Imperial College, London, Blackett Lab, Prince Consort Road, London SW7 2AZ, UK\\
	$^2$School of Physics and Astronomy, Cardiff University, Queen's Buildings, The Parade 5, CF24 3AA, Cardiff, UK\\
	$^3$INAF-Osservatorio Astronomico di Padova, Vicolo Osservatorio 5, I-35122 Padova, Italy\\
	$^4$SISSA, Via Bonomea 265, I-34136 Trieste, Italy\\
	$^5$Instituto de F\'\i sica de Cantabria (CSIC-UC), Av. los Castros s/n, 39005 Santander, Spain\\
	$^6$	Cornell University, 220 Space Sciences Building, Ithaca, NY 14853, USA\\
	$^7$Sterrenkundig Observatorium, Universiteit Gent, Krijgslaan 281 S9, B-9000 Gent, Belgium\\
	$^8$H.H. Wills Physics Laboratory, University of Bristol, Tyndall Avenue, Bristol, BS8 1TL, UK\\
	$^9$Department of Physics and Astronomy, University of California, Irvine CA 92697 USA\\
	$^{10}$Universit\"at Wien, Institut f\"ur Astrophysik, T\"urkenschanzstr. 17, 1180 Wien, Austria\\
	$^{11}$Institute of Astronomy, University of Cambridge, Madingley Road, Cambridge, CB3 0HA, United Kingdom\\
	$^{12}$SUPA\thanks{Scottish Universities Physics Alliance}, Institute for Astronomy, University of Edinburgh, Royal Observatory, Edinburgh, EH9
3HJ\\
	$^{13}$Department of Physical Sciences, The Open University, Milton Keynes, MK7 6AA, UK\\
	$^{14}$UK Astronomy Technology Centre, The Royal Observatory, Blackford Hill, Edinburgh, EH9 3HJ, UK\\
	$^{15}$Pontificia Universidad Cat\'olica de Chile, Departamento de Astronom\'ia y Astrof\'isica, Vicu\~na Mackenna 4860, Casilla 306,\\Santiago 22, Chile\\
	$^{16}$College of Graduate Studies, University of South Africa, P. O. Box 392, Unisa, 0003, South Africa\\
	$^{17}$Institut d'Astrophysique de Paris, UPMC and CNRS, UMR7095, 98bis Bd Arago, F-75014, Paris, France\\
	$^{18}$Leiden Observatory, Leiden University, PO Box 9513, NL-2300 RA Leiden, The Netherlands\\
	$^{19}$ Submillimetre Array, Hilo, HI, USA\\
	$^{20}$Instituto de Astrof\'{i}sica de Canarias, C/ V\'{i}a L\'{a}ctea, E-38200 La Laguna, Tenerife, Spain\\
    	$^{21}$Departimento de Astrof\'{i}sica, Universidad de La Laguna, E-38206, La Laguna, Tenerife, Spain\\
	$^{22}$California Institute of Technology, 1200 E. California Blvd., Pasadena, CA 91125, USA\\
$^{23}$Department of Astronomy and Department of Physics, University of Illinois, 1002 West Green Street, Urbana, IL 61801, USA\\
$^{24}$Instituto de F\'isica y Astronom\'ia, Universidad de Valpara\'iso, Avda. Gran Breta\~na 1111, Valpara\'iso, Chile\\
$^{25}$Department of Physics and Astronomy, University of Canterbury, Private Bag 4800, Christchurch 8140, New Zealand\\
$^{26}$European Southern Observatory, Karl-Schwarzschild-Str. 2, 85748 Garching-bei-Munchen, Germany\\
$^{27}$Dark Cosmology Centre, Niels Bohr Institute, University of Copenhagen, Denmark\\
$^{28}$Herschel Science Centre, European Space Astronomy Centre, ESA, 28691 Villanueva de la Ca\~nada, Spain\\
$^{29}$Centro de Radioastronom\'ia y Astrof\'isica, CRyA, UNAM, Campus Morelia, A.P. 3-72, C.P. 58089 Michoac\'an, Mexico\\
$^{30}$Departamento de F\'isica, Universidad de Oviedo, C. Calvo Sotelo s/n, 33007 Oviedo, Spain}}}
\date{}
\pagerange{\pageref{firstpage}--\pageref{lastpage}}
\pubyear{}

\begin{document}

\maketitle

\label{firstpage}

\begin{abstract}

We investigate the region around the {\em Planck}-detected z=3.26 gravitationally lensed galaxy HATLAS J114637.9-001132 (hereinafter HATLAS12-00) using both archival {\em Herschel} data from the H-ATLAS survey and using submm data obtained with both LABOCA and SCUBA2. The lensed source is found to be surrounded by a strong overdensity of both {\em Herschel}-SPIRE sources and submm sources. We detect 17 bright (S$_{870} >\sim$7 mJy) sources at $>$4$\sigma$ closer than 5 arcmin to the lensed object at 850/870$\mu$m. Ten of these sources have good cross-identifications with objects detected by {\em Herschel}-SPIRE which have redder colours than other sources in the field, with 350$\mu$m flux $>$ 250$\mu$m flux, suggesting that they lie at high redshift. Submillimeter Array (SMA) observations localise one of these companions to $\sim$1 arcsecond, allowing unambiguous cross identification with a 3.6 and 4.5 $\mu$m $\em Spitzer$ source. The optical/near-IR spectral energy distribution (SED) of this source is measured by further observations and found to be consistent with $z>2$, but incompatible with lower redshifts. We conclude that this system may be a galaxy cluster/protocluster or larger scale structure that contains a number of galaxies undergoing starbursts at the same time.

%We discuss the properties of the companions assuming they lie at z=3.26, finding that our submm-detected sources are Ultra- or Hyperluminous Infrared Galaxies (U/HLIRGs; $L_{FIR}>10^{12}$ or 10$^{13} L_{\odot}$ respectively). 

\end{abstract}


\section{Introduction}

The early history of galaxy clusters is a poorly constrained aspect of galaxy and large scale structure formation. Hierarchical clustering models predict that massive elliptical galaxies will form in the cores of what will become the most massive galaxy clusters today, but the epoch of the bulk of star formation for these galaxies is unclear. Observations of high redshift clusters (z=1---1.5) by the ISCS project (IRAC Shallow Cluster Survey, Eisenhardt et al., 2008) suggest that this is at $z>3$, and the presence of well defined red sequences of galaxies in clusters out to z$\sim$2 (Andreon et al. 2011; Gobat et al. 2011; Santos et al. 2011; Pierini et al. 2012) supports this conclusion. Theoretical models by Granato et al. (2004) suggest that forming clusters will go through a phase in which multiple members will undergo near-simultaneous massive bursts of star formation. The spectral energy distribution of these objects would be dominated by the far-IR, as is the case for local massive starbursts (eg. Clements et al., 2010). A galaxy cluster or protocluster (we use the term protocluster to indicate a structure that has yet to become virialised) going through such a formative phase would appear as a clump of dusty protospheroidal galaxies, and might be detected by observations in the far-IR and submm bands. Hints of such objects may already have been found by {\em Spitzer} (Magliocchetti et al., 2007) and SCUBA (Ivison et al., 2000; Priddey et al., 2008; Stevens et al., 2010 and references therein). A recent study by Ivison et al. (2013) has uncovered a group of HLIRGs and ULIRGs at z=2.41 thought to be the progenitor of a 10$^{14.6}M_{\odot}$ cluster. At still higher redshifts, the highest redshift protocluster currently known, at z$\sim$5.3 (Capak et al., 2011; Riechers et al. 2010), and a group of objects associated with a z=5 quasar (Husband et al., 2013) both contain at least one submm luminous object, while the best studied group of high z submm-luminous sources to date is probably the four objects associated with a source in the GOODS-North field designated GN20, all lying at z=4 (Daddi et al., 2009; Carilli et al., 2011), though this group is extended over a broad range of redshifts $\Delta z \sim 0.1$.

Large area surveys with the {\em Herschel Space Observatory} (Pilbratt et al., 2010), SCUBA2 (Dempsey et al., 2013; Holland et al., 2013), or the {\em Planck} all sky survey ({\em Planck} Collaboration, 2011; Negrello et al., 2005) are capable of finding such objects. Several studies have been undertaken combining {\em Planck} and {\em Herschel} data to identify candidate high redshift clusters while others have used {\em Herschel} data (eg Valtchanov et al., 2013; Dannerbauer et al., 2014; Rigby et al., 2014) or SCUBA2 data (Ma et al., 2015; Casey et al., 2015) alone or in combination (Noble et al., 2013). Clements et al. (2014) found a number of candidate objects in the HerMES (Oliver at al., 2012) survey at z$\sim 1 - 2$, while {\em Herschel} followup observations of {\em Planck} all sky sources ({\em Planck} Collaboration, 2015) have identified at least one z$\sim$1.7 cluster of dusty galaxies and large numbers of candidates. The sources uncovered in HerMES include two clusters lying at $z\sim 0.8$ and $z\sim 2$, confirmed by the presence of a red sequence, and two lying at $z\sim 1.1$ and $z\sim 2.3$ confirmed through photometric redshift analysis. Total star formation rates for these HerMES clusters, derived from the {\em Herschel} fluxes of their members, range from 1000 to 10000 $M_{\odot} yr^{-1}$. The highest redshift candidate cluster of dusty galaxies so far, though, has emerged from analysis of {\em Planck} sources seen in the first release of data from the H-ATLAS survey (Herranz et al., 2013). This paper describes the results of followup observations of this source.

In the next section we describe the H-ATLAS survey and the properties of the source HATLAS J114637.9-001132 (referred to in the rest of this paper as HATLAS12-00, after Fu et al., 2012) uncovered by Herranz et al. (2013) through matching the {\em Planck} Early Release Compact Source Catalog (ERCSC; {\em Planck} Collaboration, 2011) to the Phase 1 H-ATLAS maps and catalogs (Valiante et al., in prep). In Section 3 we describe the followup observations of this source, obtained using a variety of telescopes, together with the reduction and analysis of this data. The results of these observations are discussed in Section 4, while Section 5 discusses the nature of the sources found in these observations. Our conclusions are summarised in Section 6. Throughout this paper we assume a standard concordance cosmology with $H_0 = 70\, \mathrm{km}\, \mathrm{s}^{-1} \, \mathrm{Mpc}^{-1}, \Omega_M = 0.3$ and $\Omega_{\Lambda}=0.7$.

\section{H-ATLAS and the z=3.26 Lensed Source HATLAS12-00}

\subsection{The H-ATLAS Survey}

The {\em Herschel} Astrophysical Terahertz Large Area Survey (H-ATLAS, Eales et al., 2010) uses the PACS (Poglitsch et al., 2010) and SPIRE (Griffin et al., 2010) instruments on the {\em Herschel} Space Observatory (Pilbrat et al., 2010) to survey a total of 616 sq. deg. of sky, making it the largest area extragalactic survey executed by {\em Herschel}. PACS observations are made at 100 and 160$\mu$m while SPIRE covers the 250, 350 and 500$\mu$m bands. The instrumental beams have full width half-maximum (FWHM) values of 8.7, 13.1, 18.1, 25.2 and 36.9 arcseconds respectively, from 100 to 500$\mu$m, with 5$\sigma$ point source sensitivities of 132, 126, 32, 36 and 45 mJy, taking into account confusion (6-8mJy in the SPIRE bands) and instrumental noise. Map making and data reduction for the H-ATLAS survey are discussed in detail in Pascale et al. (2011) and Ibar et al. (2010), and source detection and catalogue generation are described in Rigby et al. (2011).

The fields covered by H-ATLAS include the Northern and Southern Galactic Poles (NGP and SGP) and three fields at DEC $\sim$ 0 that were selected to match the three fields of the Galaxy And Mass Assembly (GAMA, Driver et al., 2011) survey, which provides plentiful redshifts for z$<$0.5 optically selected galaxies. The phase 1 H-ATLAS data release includes the GAMA fields (Valiante et al. in prep; Bourne et al., in prep.), while further data releases will cover the larger NGP and SGP survey areas. The maps that form the basis for this release are background subtracted and convolved with a matched filter (Chapin et al. 2011) calculated to produce the maximum signal-to-noise for sources that are unresolved in the un-smoothed images. Sources are then selected from these maps using the MADX algorithm (Maddox et al., 2010; Maddox et al. in prep). Comparison of the results of this process for the match-filtered phase 1 catalogs and the un-match-filtered SDP catalogs (Rigby et al., 2011) show clear but modest improvements in both S/N and positional accuracy (see Valiante et al., in prep, for more details of the data reduction methods and comparison with previous results).

\subsection{The Lensed Source HATLAS12-00}

Herranz et al. (2013) searched for cross identifications between {\em Planck} ERCSC ({\em Planck} Collaboration, 2011) and phase 1 release H-ATLAS sources (roughly 25\% of the final 616 sq. deg.). 28 matches were found, most of which were associated with nearby quiescent galaxies detected in the optical (Herranz et al., 2013) or with extended cirrus dust structures in our own galaxy. One source, however, had no bright optical counterpart and showed no signs of cirrus contamination. The submm colours of this {\em Planck} ERCSC source are unusually red (see Fig 12 of Herranz et al. 2013) suggesting that it might lie at high redshift. Examination of the {\em Herschel} images from H-ATLAS show that it is associated with a clump of sources, several of which have red {\em Herschel} colours (ie. 350$\mu$m flux $>$ 250$\mu$m flux, see below), grouped around a single bright, optically unidentified {\em Herschel} source, designated HATLAS12-00, whose flux peaks at 350$\mu$m.

HATLAS12-00 had already been identified as a candidate gravitationally lensed galaxy as a result of its high submm flux (i.e. $F_{500}>100$mJy), red {\em Herschel} colours and the lack of a bright optical or radio counterpart (see eg. Negrello et al. 2010 for a discussion of the selection of lens candidates in H-ATLAS and other {\em Herschel} surveys). This source was therefore observed spectroscopically in the submm. A CO spectroscopic redshift of 3.26 was first suggested by Z-spec (Bradford et al., 2004) observations, then subsequently confirmed by observations by the CARMA interferometer (Leeuw et al., in prep) and the Zpectrometer instrument (Harris et al., 2007) on the Greenbank Telescope (Harris et al., 2012; see also Fu et al., 2012). Additional followup observations in the optical, near-IR, submm and other wavelengths were targeted at the lensed $z=3.26$ source and the foreground objects responsible for the lensing, resulting in detailed analyses of this lensing system by Fu et al. (2012) and Bussmann et al. (2013). Their conclusions are that the $z=3.26$ source HATLAS12-00 is subject to gravitational lensing, with a magnification of 9.6$\pm$0.5 in both the submm continuum and CO, and 16.7$\pm$0.8 in the K' band, by two foreground galaxies, one at a spectroscopically determined redshift of 1.22, and another with photometry suggesting that it lies at a similar redshift. The submm photometry of HATLAS12-00 at 890$\mu$m acquired with the Submillimeter Array (SMA) as part of this programme (Fu et al., 2012; Bussmann et al., 2013) is fully consistent with the 870$\mu$m and 850$\mu$m fluxes derived for this source from the LABOCA and SCUBA2 data to be presented here. The spectral energy distribution (SED) of the lensed source, after correcting for the lensing amplification, is well matched by the optically thick SED model for Arp220 from Rangwala et al. (2011), with a lensing-corrected far-IR luminosity of 1.2$\pm 0.2 \times 10^{13} L_{\odot}$, and an implied star formation rate of 1400 $\pm$300 $M_{\odot}$ yr$^{-1}$. In many ways the unlensed properties of this object match those of the broader population of bright submm selected galaxies first discovered by the SCUBA submm imager (see eg. Chapman et al., 2005; Clements et al., 2008; Micha{\l}owski et al., 2010). The unlensed 870$\mu$m flux of this object would be $\sim$7.7mJy.

\section{Observations of the HATLAS12-00 Field}

\subsection{LABOCA Observations}

We observed a field of diameter $\sim$600 arcsec around the position of the lensed source HATLAS12-00 with the LABOCA 870$\mu$m imager on the APEX telescope (the Atacama Pathfinder EXperiment, G{\"u}sten et al., 2006), during ESO programme 088.A-0929. APEX is a 12m diameter submm telescope situated in the Atacama desert, Chile. LABOCA is a multichannel continuum bolometer array mounted on APEX (Siringo et al, 2009) used for continuum imaging at a central wavelength of 870$\mu$m. It has 295 bolometer detectors arranged in concentric hexagons giving a total instantaneous field of view of 11.4 arcmin. The angular resolution is 18.6 arcsec (FWHM) and the channels are separated by 36 arcsec meaning that the field of view is undersampled by the array. We therefore adopted a standard spiral scanning pattern to produce a fully sampled map of the instrumental field of view. 

Observations were carried out over five nights from 21 Nov. 2011 to 25 Nov. 2011. The target field was observed for a total integration time of 20 hours, in continuous scans of 7 minutes each. Atmospheric opacities for our observations ranged from 0.158 to 0.357, with a median $\tau = 0.209$. Standard calibrators were observed during the observing run, in-between scientific scans, and included the planets Mars and Uranus, and the stars CW Leonis, B13134 and N2071IR. The radio source J1058+016 was used as a pointing calibrator throughout the whole observing run.

Data reduction used the official data reduction package BoA\footnote{Bolometer Array Analysis Software: www3.mpifr-bonn.mpg.de/div/submmtech/software/boa/boa\_main.html}. This provides a dedicated pipeline for the processing of LABOCA data and includes automated flux calibration, flatfielding, opacity correction, noise removal, and despiking of the bolometer timestreams. In addition, BoA provides a number of scripts for different reduction schemes, with preset parameters that can be modified by the user. We ran BoA using a modified version of the default script {\it reduce-map-weaksource.boa}, where we modify the low-frequency filter window to allow less aggressive filtering: this proved necessary because of the presence of the bright lensed source HATLAS12-00 at the centre of the observed field, which generated significant negative lobes if too narrow a filter was used. We found iteratively that a filtering window from 0.25 to 0.4 (compared to the default 0.3-0.35 interval) provided a satisfactory result while efficiently filtering the low-frequency noise. These values are instrumental parameters related to the details of the data reduction software, and are discussed further in section 3.8.4 of the BoA manual (Schuller et al., 2010). The scans are then variance-weighted and coadded to provide final maps of flux density and noise. The final reduced map provides a useful field of $\sim$11 $\times$ 9 arcmin in size with typical 1$\sigma$ noise ranging from 1.4mJy to 2.5mJy.

Individual sources are extracted from the map using a source-finding algorithm which identifies peaks in a smoothed signal-to-noise map, produced by convolving the flux density map, weighted by the inverse of the variance, with the 18.6 arcsec full width half maximum (FWHM) beam. Peaks with a signal-to-noise ratio of at least 2.5 were selected as candidate sources, their flux (error) was then calculated as the value of the beam-convolved flux density (noise) map at the position of the peak. We detect seven peaks above a 4$\sigma$ threshold, four of which have S/N $>$5 (see Table 1). One of these, as expected, is the lensed source HATLAS12-00.

\begin{table*}
\begin{tabular}{cccccc}\hline
Name&RA&Dec&F870&S/N&Comments\\ 
&Deg J2000& Deg J2000& mJy\\ \hline
LABOCA\_1  &176.658858      &-0.191142     &79.4  $\pm$ 2.5     &  32.1      &    5" from HATLAS12-00, lensed source\\
LABOCA\_2  &176.674044 & -0.226574  &14.8  $\pm$ 1.9   &  7.9    &    136" from lens; 5" from HATLASJ114641.4-001332\\
LABOCA\_3  &176.679105  & -0.137993  &10.3  $\pm$ 1.6   &  6.6    &   207" from lens\\
LABOCA\_4  &176.598117  & -0.183549  &10.8  $\pm$ 2.1  &   5.1   &    220" from lens; 2" from HATLASJ114623.5-001058\\
LABOCA\_5  &176.684167 &  -0.145586  &7.3  $\pm$ 1.5   &  4.9   &     190" from lens; 6" from HATLASJ114644.6-000840\\
LABOCA\_6  &176.608240 &  -0.155709 & 8.6  $\pm$ 2.1  &   4.2   &     223" from lens\\
LABOCA\_7  &176.742378  & -0.196203 & 9.4 $\pm$  2.3  &   4.0    &    301" from lens\\ \hline
\end{tabular}
\caption{Sources detected with S/N $>$4 by the LABOCA observations, giving the position, LABOCA flux and S/N at 870$\mu$m as well as the angular distance from the lensed source HATLAS12-00. LABOCA\_1 can readily be identified with the lensed source itself.}
\end{table*}

\subsection{SCUBA2 Observations}

A field of diameter $\sim$ 700 arcsec around the position of HATLAS12-00 was observed with the SCUBA2 850 \& 450$\mu$m imager (Holland et al., 2013) on the James Clerk Maxwell Telescope (JCMT) on Mauna Kea as part of a programme to observe candidate clusters of dusty galaxies detected through cross matching {\em Planck} and {\em Herschel} sources. SCUBA2 provides an angular resolution of 14.5 arcsec (FWHM) at 850$\mu$m.

Observations for the programme were carried out over a five night period from 8 April 2013 to 12 April 2013, but the observations of HATLAS12-00 discussed here were taken only on 8th and 11th April. Since the target for these observations was the immediate region around the lensed source, we used the DAISY observing mode, which scans the detector arrays in a `spirograph' pattern around the targeted position (in this case HATLAS12-00). This optimises the sensitivity in the central regions of the field, but provides reasonable sensitivity over a larger field. HATLAS12-00 was observed for four separate observing blocks, each of 30 minutes duration, for a total integration time of two hours. Three of these blocks were observed on 8th April, and one on 11th April. Observing conditions were good (grade 2 or upper grade 3, corresponding to 1 - 2mm of precipitable water vapour) for all of these observations. Standard observatory calibration observations were taken on both nights.

Data reduction used the standard SMURF package (Chapin et al. 2013) following the procedures outlined in the SCUBA2 Data Reduction Cookbook (Thomas, 2012). The separate scans were reduced using the dynamic iterative mapmaker {\em makemap}. These were then combined using a noise-weighted optimal method using the {\em MOSAIC\_JCMT\_IMAGES} recipe in the PICARD environment. Since we are interested in unresolved, faint point sources, we used the {\em SCUBA2\_MATCHED\_FILTER} recipe, which removed large angular scale variations in this final map by calculating a smoothed map with a 30 arcsec kernel, subtracting this, and then convolving the map with the 850$\mu$m beam. The resulting image then had the higher-noise edge regions cropped, limiting it to a circular field with diameter 700 arcsec. Calibration from observed detector power to detected flux densities used the standard observatory determined flux calibration factor of 537 Jy/pW/beam at 850$\mu$m. Noise in our final image ranged from 1$\sigma$=1.3mJy in the centre to 1$\sigma$=2.7mJy at the edges.

Individual sources were extracted from the map by identifying peaks in the S/N map calculated by dividing the final beam convolved map by the noise map. Applying a 4$\sigma$ threshold for detection, we find a total of fourteen 850$\mu$m sources. Six of these, including the lensed source, have a S/N $>$5 (see Table 2).

Observing conditions were not good enough for the 450$\mu$m data simultaneously taken with the 850$\mu$m observations to be useful.

\begin{table*}
\begin{tabular}{cccccc}\hline
Name&RA&Dec&F850 &S/N&Comments\\
&Deg J2000&Deg J2000&mJy\\ \hline
SCUBA2\_1              & 176.6068780   &   -0.2351671   &    10.6    $\pm$2.1     &    5.0     &   Out of LABOCA field; 243" from lens; 4" from HATLAS\_78812\\
SCUBA2\_2          &      176.6161110   &   -0.1266438   &    9.6     $\pm$2.3      &   4.2    &  281" from lens\\
SCUBA2\_3          &      176.6811353   &   -0.1372072    &   10.5  $\pm$2.2       &    4.7   &     7.8" from LABOCA\_3; 213" from lens\\
SCUBA2\_4          &      176.6749951   &   -0.1491695    &   8.6   $\pm$2.0       &    4.2     &   165" from lens\\
SCUBA2\_5          &      176.6860210   &   -0.1568639  &     10.5   $\pm$2.0     &     5.3    &      160" from lens\\
SCUBA2\_6          &      176.5883331   &   -0.1638748   &    9.9   $\pm$2.5       &    4.0    &    273" from lens\\
SCUBA2\_7         &       176.6155570   &   -0.1716642   &    8.0   $\pm$2.0       &    4.1    &    172" from lens\\
SCUBA2\_8        &        176.6477698   &   -0.1805539   &    7.0    $\pm$1.4      &    5.0     &   57" from lens; 4" from HATLASJ114635.1-001048\\
SCUBA2\_9        &        176.6173862  &    -0.1857557    &   10.6   $\pm$1.8     &     6.0     &   151" from lens; 8" from HATLASJ114628.6-001114\\
SCUBA2\_10      &         176.6587577  &    -0.1919366   &    80.0  $\pm$1.3     &      59.7   &     2.9" from LABOCA\_1; 0" from lens, lensed source\\
SCUBA2\_11      &         176.6524809  &    -0.2122232   &    6.4    $\pm$1.5     &     4.4     &   76" from lens\\
SCUBA2\_12      &         176.6415322   &   -0.2196134   &    7.5   $\pm$1.6      &     4.7      &  117" from lens\\
SCUBA2\_13       &        176.6738450   &   -0.2257691     &  10.0  $\pm$1.7    &      5.8    &    3.0" from LABOCA\_2; 133" from lens\\ 
SCUBA2\_14       &        176.6661112  &    -0.2338934  &     7.3   $\pm$1.8       &    4.0     &   153" from lens\\ \hline
\end{tabular}
\caption{Sources detected with S/N $>$4 by the SCUBA2 observations, giving the position, SCUBA2 flux and S/N at 850$\mu$m as well as the angular distance from the lensed source HATLAS12-00 and from any nearby LABOCA sources. As with LABOCA\_1, SCUBA2\_10 can readily be identified with the lensed source itself.}
\end{table*}

\subsection{{\em Spitzer} Observations}

{\em Spitzer}-IRAC 3.6 and 4.5 $\mu$m staring mode observations of the extended region covering the clump of sub-mm galaxies around HATLAS12-00 were carried out during the warm mission.  In staring mode, simultaneous observations at 3.6 and 4.5 $\mu$m are offset from each other by 6.8 arcminutes. Due to the offset, $\sim$0.05 deg$^2$ of the targeted area has imaging data at both 3.6 and 4.5 $\mu$m, and the remainder split between 3.6 and 4.5 $\mu$m coverage. The total exposure time is 1600 seconds per pixel in each filter with the four-point dithered map taking over four hours to complete.  Data reduction and mosaicking was performed on the Corrected Basic Calibrated Data (cBCD) provided by the {\em Spitzer} Science Center using MOPEX (MOsaicker and Point source EXtractor; Makovoz \& Marleau, 2005). The final mosaicked images have 0.6 $\times$ 0.6 arcsecond pixels and the FWHM of the IRAC point spread function (PSF) at 3.6 and 4.5$\mu$m is 1.66 and 1.72 arcseconds, respectively. Source detection is performed with SEXTRACTOR (Bertin \& Arnouts 1996) and detected sources are required to comprise at least three contiguous pixels with fluxes at least 1.5$\sigma$ above the local background. Photometry was measured in 3.8 arcsecond diameter apertures with the APPHOT task in IRAF. The advantage of APPHOT is that the photometry is measured in fixed apertures at specified source positions. The measured aperture photometry was corrected to ÒtotalÓ fluxes, assuming point-source profiles, and using the calibration derived by the SWIRE team for IRAC data with multiplicative correction factors of 1.36 and 1.40 at 3.6 and 4.5 $\mu$m, respectively (Surace et al. 2005).

\subsection{Submillimetre Array Observations}

Four regions around HATLAS12-00 were observed in track sharing mode at the Submillimeter Array on Mauna Kea in Compact configuration on 2013 April 7 UTC and 2013 May 16 UTC. The SIS receivers were tuned to an LO frequency of 270.0 GHz. Both
observations were made in good conditions, with
$\tau_{225}\sim$0.07 and 10-25\% humidity.  An additional track in SubCompact configuration was obtained on 2013 May 13 UTC in slightly worse conditions ($\tau_{225}\sim$0.07 and only 6 antennas) but was combined with the other two tracks for a slight improvement in SNR and a final beam of about 2.7 x 2.3 arcsec. The targets for these observations were sources LABOCA\_2, LABOCA\_3, LABOCA\_4 and LABOCA\_5, with predicted 270GHz fluxes of  8, 4, 6 and 4mJy respectively.

The SMA data were calibrated using the MIR software package developed
at Caltech and modified for the SMA. Gain calibration was performed
using the nearby quasars 3c279 and 3c273. Absolute flux calibration was
performed using realtime measurements of the system temperatures with
observations of Titan to set the scale, and bandpass calibration was
done using 3c279 and 3c84.

The data were imaged using MIRIAD (Sault et al., 1995), and further
analysis carried out using the KARMA software package. The noise in the final co-added maps ranged from 1-1.3 mJy. A secure source, $>4\sigma$, was only detected in the map of LABOCA\_2 with a flux of 4.2 $\pm$ 1 mJy. It is also detected in both the individual (uncombined) Compact and SubCompact maps. The brightest sources in the SMA maps associated with the positions of the other LABOCA sources were 3.8$\pm$1.3, 4.3$\pm$1.3 and 3.1$\pm$1.1 mJy for LABOCA\_3, LABOCA\_4 and LABOCA\_5 respectively, which are compatible with the fluxes predicted for the sources.

%Small regions around four of the sources detected by LABOCA and/or SCUBA2 were observed by the Submillimetre Array at 270 GHz in Compact configuration. Data were taken on 7th April and 16 May 2013. The sources observed were LABOCA\_2,3,4 and 5, details of which can be found in Table 1. The sources were observed using track sharing, with the array switching between each source throughout the track. These observations took a total of 19 hours of observations during two tracks. 3C279, 3C273 and 1058+015 were used as gain calibrators for these observations, while MWC349a and Titan were used as flux calibrators. 3C279 and 2015+371 were used as bandpass calibrators. The data were reduced using standard routines in MIRIAD (Sault et al., 1995) and KARMA. The noise in the final coadded maps ranged from 1 to 1.3mJy. A secure source was only detected in the map of LABOCA\_2  with a flux of 4.2 $\pm$1 mJy. 

\subsection{Optical and Near-Infrared Observations}

Near-IR observations in the J and K' bands were taken of a $\sim$150 x 150 arcsecond region around the position of the SMA-identified source LABOCA\_2 (see above) using the ISAAC instrument on the VLT. J and K' observations were taken on the 19th April 2013 and 12th March 2013 respectively. Observing conditions for both these observations were good with seeing measured at 0.7 and 0.6 arcseconds respectively. Total on-source integration times were 3240s at J and 1536s at K'. These observations were conducted in service mode. The data were reduced using the standard ESO processing pipeline, and the images were astrometrically and photometrically calibrated using cross identifications with 2MASS sources.

Optical observations using the ACAM imager on the William Herschel Telescope were used to provide a deep, white light (ie. filter free) image of the field around the HATLAS12-00 lens. These observations are discussed in detail in Fu et al. (2012) and reach a 5$\sigma$ detection limit comparable to $i_{AB} = 24.6$.

\section{Observational Results}

The SCUBA2 850$\mu$m image is shown in Fig 1, with the LABOCA S/N contours overlaid. In Fig 2 we present a three colour SPIRE image of this field (blue for 250$\mu$m, green for 350$\mu$m and red for 500$\mu$m) with the positions of the SCUBA2 and LABOCA detected sources indicated. The lensed source is clearly detected, and a total of 17 other LABOCA and/or SCUBA2 sources are also found with S/N $> 4\sigma$. 

\begin{figure*}
\vspace{-6cm}
\epsfig{file=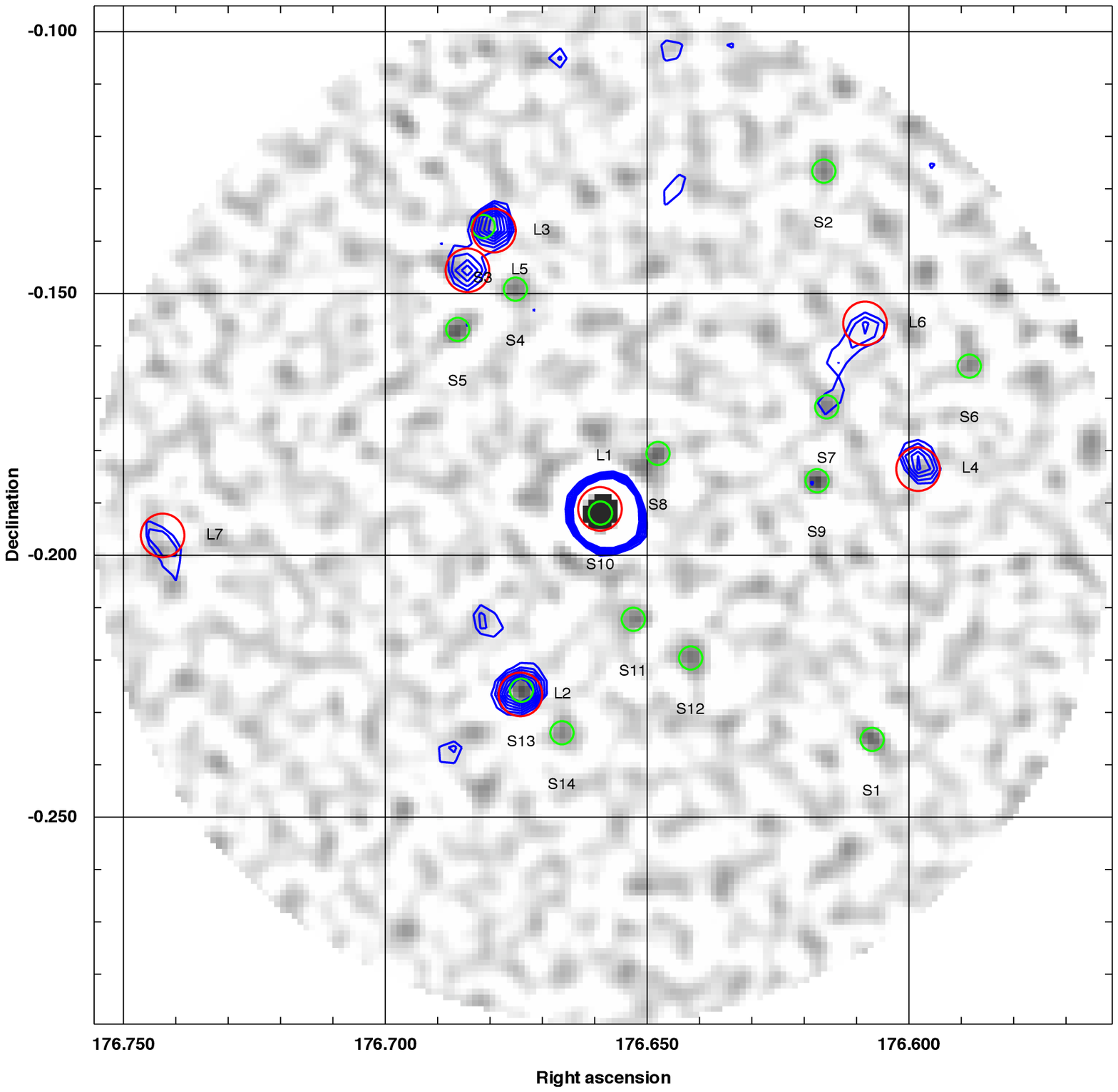, angle=0, width=20cm}
\vspace{-5cm}
\caption{SCUBA2 850$\mu$m image of the clump, LABOCA signal-to-noise contours overplotted in blue, positions of $>4\sigma$ LABOCA sources (red) and SCUBA2 sources (green) indicated by circles. LABOCA S/N contours start at 3$\sigma$ and rise to 10$\sigma$ in steps of 1$\sigma$. Source position circle size is matched to the beam FWHM of the appropriate instrument. Source names shown in black as L1, S1 etc.. For SCUBA2 sources the name lies below the relevant source, for LABOCA, with the exception of the lensed source, they lie to the right. For the lensed source it lies above. Coordinates are given in degrees J2000.}
\end{figure*}

\subsection{Cross Identification with SPIRE Sources}

To obtain cross identifications and SPIRE flux measurements for the sources found by LABOCA and SCUBA2 we examine both the SPIRE catalogs produced by the H-ATLAS team and the SPIRE maps themselves. We first compare the 850/870$\mu$m sources identified by SCUBA2 and LABOCA to the officially released H-ATLAS source catalog (Valiante et al., in prep). This catalog includes sources detected at $>5\sigma$ significance in at least one SPIRE band. This comparison yields six clear identifications, including the lensed source, with separations between the SPIRE and submm sources of 9 arcseconds or less (see Table 3), consistent with the expected combined positional uncertainties of these datasets of $\sim$10 arcseconds. The fluxes and H-ATLAS names for these sources are reported in Tables 1, 2 and 3 as appropriate. Given the density of 850$\mu$m sources on the sky and our matching radius we expect only $\sim$0.07 false positive identifications of a SPIRE source with a submm source over the entire catalog of 18 submm sources. If we instead use the enhanced submm source density over the field that we find in this region (see Section 4.3) then the number of false positives predicted rises to 0.3. We can thus consider these identifications reliable.

We also compare the LABOCA and SCUBA2 source lists to a supplementary H-ATLAS catalog generated using the same MADX algorithm (Maddox et al., 2010; Maddox et al. in prep) as the main catalog but extending down to 3$\sigma$ significance detections rather than the 5$\sigma$ used for the main catalog. Cross matching with this extended catalog provides a further five SPIRE identifications for our LABOCA and/or SCUBA2 sources with separations of 9 arcseconds or less. Identifications with this supplementary catalog are indicated in Tables 1, 2 and 3 as appropriate. One further plausible cross identification, with a separation of 10.8 arcseconds is also found, for SCUAB2\_2. For these supplementary identifications and for submm sources not directly associated with a catalogued SPIRE source, we estimate their SPIRE fluxes by extracting the fluxes at the LABOCA or SCUBA2 positions from the background-subtracted match-filtered SPIRE maps provided by H-ATLAS. Source fluxes from SPIRE, LABOCA and/or SCUBA2 are shown in Table 3.

One source, SCUBA2\_3 (which is cross identified with LABOCA\_3 and an H-ATLAS supplementary source) lies fairly close to a second SPIRE source (LABOCA\_5) with a separation of $\sim$17". This is comparable to the size of the SPIRE 250$\mu$m beam and smaller than the 350 or 500$\mu$m beams. The immediate neighbourhood of this source also encompasses the sources SCUBA2\_5 and SCUBA2\_4, all within an area of about 1 sq. arcmin. The source density in this region is thus substantially higher than the usual one source per 20 beams criterion for confusion (Hewish, 1961). We might thus expect problems with flux estimation in this region. To address this issue we extract far-IR fluxes from the H-ATLAS maps at 250, 350 and 500$\mu$m for this confused region using the positions of the submm detected objects, as noted above, as priors. We also perform a simultaneous fit to the maps using four gaussians, with FWHMs corresponding to the instrumental beams at the appropriate wavelength, and with positions corresponding to those of the submm detected objects, to a section of the images around these sources using the image fitting tool Imfit (Erwin, in prep)\footnote{http://www.mpe.mpg.de/$\sim$erwin/code/imfit/}. The gaussians are fitted simultaneously to, in effect, deblend the sources. We found that this deblending procedure produces fluxes which are consistent with those extracted directly from the maps. The deblended fluxes are those given in Table 3.

\subsection{Comparison of LABOCA and SCUBA2 Source Lists}

A combination of both the LABOCA and SCUBA2 source catalogs provides a list of 18 distinct sources, with three sources, including the lens HATLAS12-00,  appearing in both, and a total of twelve sources cross identified between one (or both) of the submm source lists and the H-ATLAS SPIRE catalog. This raises the question of the reality of those submm sources detected by one set of observations and not the other. This issue is complicated by the fact
that the noise in the two submm data sets is not uniform, by confusion, by different beamsizes, and by the differences in the data reduction methods used by the two bolometer array instruments. To address this further, we have extracted the SPIRE fluxes of each of the submm sources as described above. Where sources are not explicitly detected by both submm instruments, the submm fluxes from the complementary submm map are also extracted. These are given in Table 3.

We assess the reliability of our submm detected sources by comparing the S/N with which they are detected in the five bands observed by the three instruments. Sources detected by SCUBA2 and/or LABOCA are considered good if they are detected in a total of at least three bands from SPIRE (250, 350 and 500$\mu$m), LABOCA and SCUBA2. In this context, detection in a SPIRE band is considered to be a S/N $>$3, whether this comes from the H-ATLAS catalogue or is measured from the H-ATLAS maps. Marginal sources are those detected by only one complementary band, and unreliable sources are those not detected in any complementary passbands. On this basis we have 10 good detections, including the lensed source, 5 marginal detections, and 3 unreliable detections.

As can be seen from the table notes, three of the SCUBA2 sources are found in parts of the LABOCA map where there are problems with the data reduction. One of these, SCUBA2\_8, is close to the bright lensed source HATLAS12-00, and is affected by negative ringing in the LABOCA map production process. Nevertheless, it is clearly detected by SPIRE, so its reality is not in doubt. Two other SCUBA2 sources lie in a region where there is a large-scale gradient in the LABOCA map. We attempt to remove the effects of this gradient by smoothing the LABOCA map with a Gaussian with a 1$\sigma$ of 1 arcminute, subtracting this from the LABOCA map, and then searching for flux at the positions of these sources. In neither case does this suggest the presence of a source detected by LABOCA at $>$3$\sigma$, but the presence of this gradient suggests that the LABOCA data in this area may be of a lower quality than the gaussian noise suggests. The lack of LABOCA confirmation for SCUBA2\_12, a source clearly detected by SPIRE at $>3 \sigma$ in all three bands is thus unlikely to be a problem.

%In general, the least reliable of our sources are those with the lowest S/N detections in the submm data, close to our 4$\sigma$ threshold. This is not unexpected. However, the detection criterion we are using here is somewhat more conservative than that used for many of the extant or ongoing large scale surveys using either SCUBA2 or LABOCA. The SCUBA2 Cosmology Legacy Survey (Geach et al., in prep), for example, uses a 3.5$\sigma$ threshold for detection while LESS (Weiss et al., 2009) used a 3.7$\sigma$ threshold. The unreliability of our $\sim$4$\sigma$ sources would suggest the need for caution when dealing with the least significant sources in these other submm surveys. We note that the reality, or otherwise, of the lowest S/N sources in such surveys is a different issue to that of flux boosting (also known as Eddington bias), where the flux of a real source is biassed upwards by the underlying confusion noise. 

\begin{table*}
\begin{tabular}{cccccccc}\hline
Name&F250&F350&F500&F$_{LABOCA}$&F$_{SCUBA2}$&Comments&Status\\
&(mJy)&(mJy)&(mJy)&(mJy)&(mJy)\\ \hline
LABOCA\_1  &290$\pm$6&356$\pm$7&295$\pm$8& 79.4  $\pm$ 2.5 & 80$\pm$1.4 & HATLAS12-00, also SCUBA2\_10&Good\\
LABOCA\_2  &51$\pm$6 & 54 $\pm$7& 50$\pm$8 &14.8  $\pm$ 1.9 & 10$\pm$1.7    & HATLASJ114641.4-001332, also SCUBA2\_13&Good\\
LABOCA\_3$^\&$  &24$\pm$6&37$\pm$7&38$\pm$8&10.3  $\pm$ 1.6 &10.5$\pm$2.2&HATLAS supplementary;  SCUBA2\_3&Good\\
LABOCA\_4  &40$\pm$6&42$\pm$7&14$\pm$8 &10.8  $\pm$ 2.1  &6.4$\pm$2.1  &   HATLASJ114623.5-001058&Good\\
LABOCA\_5$^\&$  &26$\pm$6&47$\pm$7&46$\pm$8&7.3  $\pm$ 1.5 &7.3$\pm$2.2$^*$  & HATLASJ114644.6-000840&Good\\
LABOCA\_6$^\#$  &11$\pm$6&16$\pm$7&7$\pm$8& 8.6  $\pm$ 2.1  & 0.6$\pm$2.3 &  &Unreliable\\
LABOCA\_7$^\#$  &10$\pm$6&10$\pm$7&4$\pm$8&9.4 $\pm$  2.3&8.2$\pm$2.7$^!$  &  & Marginal\\ 
SCUBA2\_1$^\#$ &12$\pm$6&18$\pm$7&18$\pm$8&&          10.6 $\pm$2.1&HATLAS supplementary; &Marginal\\
&&&&&&Outside LABOCA field; \\
SCUBA2\_2$^\#$  &18$\pm$6&8$\pm$7&-2$\pm$8& 0$\pm$2&   9.6     $\pm$2.3&Possible HATLAS supplementary ID&Marginal\\
SCUBA2\_4$^\&$  &16$\pm$6&18$\pm$7&26$\pm$8&  3.3$\pm$1.5&   8.6   $\pm$2.0&HATLAS supplementary&Marginal\\
SCUBA2\_5$^\&$  &18$\pm$6&23$\pm$7&27$\pm$8& 4.3$\pm$1.4  &10.5   $\pm$2.0&HATLAS supplementary&Good\\
SCUBA2\_6$^\#$  &3$\pm$6&5$\pm$7&-6$\pm$8&   0$\pm$2.5 &    9.9   $\pm$2.5&&Unreliable\\
SCUBA2\_7$^\#$  &21$\pm$6&23$\pm$7&15$\pm$8&7$\pm$2.0&    8.0   $\pm$2.0 &HATLAS supplementary&Good\\
SCUBA2\_8  &55$\pm$6&65$\pm$7&51$\pm$8&-2.5$\pm$2.4$^\%$ &    7.0    $\pm$1.4      &HATLASJ114635.1-001048;&Good\\
&&&&&& Poor LABOCA data\\
SCUBA2\_9  &56$\pm$6&60$\pm$7&31$\pm$8&6$\pm$2      & 10.6   $\pm$1.8     &  HATLASJ114628.6-001114&Good\\
SCUBA2\_11$^\#$ &5$\pm$6&11$\pm$7&24$\pm$8&0$\pm$2.3    &    6.4    $\pm$1.5     &Gradient affecting LABOCA flux&Marginal\\
SCUBA2\_12$^\#$  &22$\pm$6&30$\pm$7&24$\pm$8&0$\pm$2.2    & 7.5   $\pm$1.6      &HATLAS supplementary; &Good\\
&&&&&&Gradient affecting LABOCA flux; \\
SCUBA2\_14$^\#$ &4$\pm$6&8$\pm$7&14$\pm$8& 0$\pm$1.8     & 7.3   $\pm$1.8       & &Unreliable  \\ \hline

\end{tabular}
\caption{Comparison of fluxes from LABOCA, SCUBA2 and SPIRE for all submm detected sources with cross identifications in the relevant catalogues noted. The errors quoted for SPIRE fluxes include the confusion noise. Notes: $^\#$ SPIRE fluxes measured from the map. $^*$ SCUBA2 flux not from the exact position of the LABOCA source, but from a plausible 3.4$\sigma$ SCUBA2 detection $\sim$4" from the position of the SPIRE source. $^!$ SCUBA2 flux not from the exact position of the LABOCA source, but from a possible 3.0$\sigma$ SCUBA2 detection $\sim$10" from the LABOCA source. $^\%$ LABOCA flux affected by negative ringing from the bright lensed source. $^\&$ SPIRE fluxes extracted from map by deblending using imfit.}
\end{table*}

\subsection{Source Overdensities Around HATLAS12-00}

Our observations of the region around HATLAS12-00 have uncovered a number of potentially associated sources at both submm wavelengths, as found by SCUBA2 and LABOCA, and in the {\em Herschel} data themselves. These results are not dissimilar to the submm source overdensities uncovered by Stevens et al. (2010) around quasars, the {\em Herschel} source overdensities found by Rigby et al. (2014) around some high redshift radio galaxies, the overdensity of {\em Herschel} and LABOCA sources associated with the Spiderweb galaxy (Dannerbauer et al., 2014), or the {\em Herschel} and SCUBA2 overdensity found behind the RCS 231953+00 supercluster by Noble et al. (2013). We here quantitatively asses the source overdensities of both submm and {\em Herschel} sources around HATLAS12-00.

The analysis here, of both SPIRE and submm source densities, compares the source densities associated with our sources to the density of individual field sources at all redshifts down to the relevant flux limits. The possibility that our clumps might be the result of line of sight alignments of several weak, physically unassociated, overdensities, as has been seen, for example, in the optical for  Abell clusters (Sutherland \& Efstathiou, 1991), is discussed in Section 5.2.

\subsubsection{SPIRE Source Density}

%An alternative assessment of the presence of an overdensity associated with HATLAS12-00 can be obtained by considering the distribution of SPIRE sources with properties similar to the counterparts of our submm sources ie. F350$>$40mJy and F350$>$F250, consistent with a redshift matching that of the lens. Examination of the H-ATLAS catalogue for the GAMA 12 field finds a 10$\sigma$ overdensity of such sources in a region of size $\sim$13 arcmin $\times$ 13 arcmin close to the position of HATLAS12-00. This is the most significant overdensity of such sources in the whole GAMA12 Phase 1 field, which covers roughly 25 sq. deg.

To determine the level of SPIRE source overdensity around HATLAS12-00 we compare the SPIRE source density around HATLAS12-00 with that in  $10,000$ randomly selected positions in the H-ATLAS GAMA12 field, which covers approximately 25 square degrees (Valiante et al. in prep). For a SPIRE source to be considered real in this comparison we require a $5\sigma$ detection in at least one SPIRE band. The furthest good companion source to HATLAS12-00 lies 220 arcseconds away so we initially assess the source density within a 220 arcscond radius region for our 10,000 comparison fields. Fourteen sources are detected by SPIRE at $>5\sigma$ in the 220 arcseconds around HATLAS12-00. Only 5 of our 10,000 comparison positions have a similar or greater source density. The HATLAS12-00 region can thus be regarded as having a probability of only $5\times10^{-4}$ of arising by chance. We also look at the level of overdensity in this field on a range of scales. We find that within 2, 2.5, 3, 4, 5, and 10 arcminute radii the H12-00 region is overdense at the $3\times10^{-3}$, $<\times10^{-4}$ (ie. no other region in our 10000 comnparison fields has a similar SPIRE source density), $3\times10^{-4}$, $9\times10^{-3}$, $1\times10^{-2}$ and $2\times10^{-2}$ levels. This confirms the presence of a significant overdensity of SPIRE sources on scales matching the Planck beam ($\sim$2.5 arcmin radius).

If we restrict our analysis to sources detected at 350$\mu$m rather than those detected in any SPIRE band then regions within 2, 2.5, 3, 4, 5 and 10 arcminutes of H12-00 are overdense at the $7\times10^{-3}$, $<  10^{-4}$ (ie. no other region in our 10000 comparison fields has a similar 350$\mu$m source density), $<  10^{-4}$, $8\times10^{-4}$, $3\times10^{-4}$ and $3\times10^{-4}$ levels, suggesting the possible presence of an overdensity on larger scales than the Planck beam, up to a radius of $\sim$10 arcmin. 
While this would certainly be a large structure in terms of physical size, amounting to about 10Mpc in diameter if it lies at the same redshift as HATLAS12-00, it is comparable to the scale of the z$\sim$3 candidate supercluster identified by Noble et al. (2013), through a combination of SPIRE and SCUBA2 data, and a number of other known protoclusters (Casey, 2016). Examination of the progenitors of $z=0$ massive clusters in the Millennium Simulation (Diener et al., 2015) suggests that they can be as large as 20Mpc across at $z\sim3$, consistent with the sizes of these overdensities identified in the far-IR/submm.

\subsubsection{Submm Source Density}

At submm wavelengths, 850 and 870$\mu$m, we do not have anything comparable to the large GAMA 12 field observed in the same way as the region around HATLAS12-00 to use as a direct comparison.
We thus apply two alternative approaches to the determination of any overdensity of submm sources associated with HATLAS12-00. The first uses the density of submm sources in the field determined by blank field surveys. The largest of these currently available by far is the SCUBA2 Cosmology Legacy Survey (CLS, Geach et al., 2016) which covers a total area of about 5 square degrees in a variety of different fields.
%
%
%Source counts from Chen et al. (2013) using SCUBA2 supersede those from the LESS survey (Wei\ss~et al., 2009) and the SHADES survey (Coppin et al., 2007) and represent the current best available data sets for this purpose, though the results from the SCUBA2 Legacy Survey (CLS; Geach et al., in prep) will supersede these once they are fully analysed.
For sources classified as good we find 9 submm companions to the lens, the maximum separation from which is 220 arcseconds. Where possible, we combine the 850 and 870 $\mu$m fluxes for these sources to get a combined noise weighted average flux. The faintest combined submm flux for a good source is 6.3 mJy for SCUBA2\_5. The fluxes for all the other sources are $\geq$7mJy. Conservatively, since the sources are not spread uniformly through this area, we use this to define a circular region of radius 220 arcseconds within which the observed source density can be compared to the field source density at the same flux limits. The SCUBA2 CLS (Geach et al., 2016) measure counts for their survey to be 130$^{+6.7}_{-6.4}$ deg$^{-2}$ at 6.5 mJy and 
208$^{+8.5}_{-8.2}$ deg$^{-2}$ at 5.5 mJy. Linear interpolation between these two values gives an estimate for the counts at 6.3 mJy to be 145$^{+7}_{-6.8}$ deg$^{-2}$. This corresponds to 1.7$^{+0.1}_{-0.08}$ sources that would be found in this area at the field density, assuming they are distributed randomly. We find 9 sources, so the probability of the observed number of companions arising randomly is then 6 $\times 10^{-5}$.

% The number of submm sources , is 3.6 using the Chen et al. counts. The probability of the observed number of companions arising randomly is then 0.0076.

%A less conservative analysis would come if we were to consider all the submm sources, good, marginal and unreliable, but this is likely an over-estimate of the source density because of the unreliability of these sources.
%
% the furthest of which is 301 arcseconds from the lens, and thus use a circular region with radius 301 arcseconds for comparison, we find 17 companions. The field counts would give an expectation of 6.7 sources in such a region. The probability of the observed number of companions arising at random is then 0.00038. 

Real submm sources are, however, not distributed purely randomly, though the clustering between them in the linear regime is unlikely to be sufficient to substantially reduce the significance of the source overdensity we have found compared to the field (see eg. Noble et al., 2013). Nevertheless, to be certain that this is not a problem we use a similar method to that used in Section 4.3.1 for SPIRE sources as a second approach to determine the significance of the submm source overdensity. For this we use the SCUBA2 CLS observations of the COSMOS field, covering $\sim$0.65 sq. deg., and the UDS field, covering $\sim$1 sq. deg, as comparisons since these are the two largest contiguous fields in the survey, and, once again, compare the number of good sources found in regions of a specific size around HATLAS12-00 to 10,000 randomly selected reference positions in the field. We do this for regions with 2, 2.5, 3 and 4 arcminutes of H12-00. We find that within 2, 2.5, 3, and 4 arcminute radii the H12-00 region is overdense at the $2.5\times10^{-2}$, $1.8\times10^{-2}$, $1.5\times10^{-3}$,  and $6.5\times10^{-4}$ levels. This confirms the presence of a significant overdensity of submm sources.
It should be noted that the comparison fields for this analysis are small in size compared to the 25 sq. deg. comparison field used for SPIRE. There is thus a possibility for some cosmic variance effects, but they are unlikely to substantially reduce the significance of this overdensity. These will also be somewhat ameliorated by combining the results, as we have done above, for the UDS and COSMOS fields.

We also compared the density of good sources around HATLAS12-00 to the SCUBA2 source overdensity discovered by Noble et al. (2013), suggested, on the basis of {\em Herschel} to SCUBA2 colours, to be a protocluster at $z \approx$ 3.
Both SCUBA2 maps have roughly equal 1 $\sigma$ sensitivities of $\sim$ 1.5mJy.
Around HATLAS12-00, we detect a total of 9 good sources over an area covering approximately 100 square arcminutes.
This corresponds to a source density of $\sim$ 324 sources per square degree.
This compares to the 16 sources detected by Noble et al. over approximately 470 square arcminutes, which gives a surface density of $\sim$ 122 sources per square degree. 
From this, we infer that HATLAS12-00 is one of the strongest overdensities of 850$\mu$m sources currently known, with a surface density $>$ 2.5$\times$ that of the structure detected by Noble et al. (2013).

\subsection{{\em Herschel}-submm Colours}

The use of SPIRE colour-colour diagrams to constrain redshifts is well established (eg. Schulz et al., 2010; Amblard et al., 2010; Negrello et al., 2010; Wardlow et al., 2013; Herranz et al., 2013; Noble et al., 2013; Clements et al., 2014) though their interpretation can be uncertain. The observed colours of sources are compared to the colours, as a function of redshift, of the spectral energy distributions (SEDs) of known low redshift template objects, such as the starburst galaxy M82 or the nearest ULIRG, Arp220. We have 10 good sources associated with HATLAS12-00, including the lens, for which reasonable quality SPIRE and submm data are available. The SPIRE colours for these sources are shown in Fig. 3, together with the SPIRE colours for the 32 SPIRE-detected sources with S/N $>$ 2 in all bands that lie in the submm fields that were not detected at $>4\sigma$ in the submm. We extend this to the submm in Fig. 4, plotting SPIRE 250-to-350$\mu$m colour against 350-to-850$\mu$m.  We have also included the SED of a cold spiral galaxy to illustrate the effects of the dust temperature-redshift degeneracy, whereby a source with colder dust can have similar far-IR-submm colours to those of a higher redshift higher dust temperature object. The dust temperature for this cold galaxy template is T=20K, and the SED is comparable to that of the local spiral M100 (Cortese et al., 2014).

\begin{figure*}
\epsfig{file=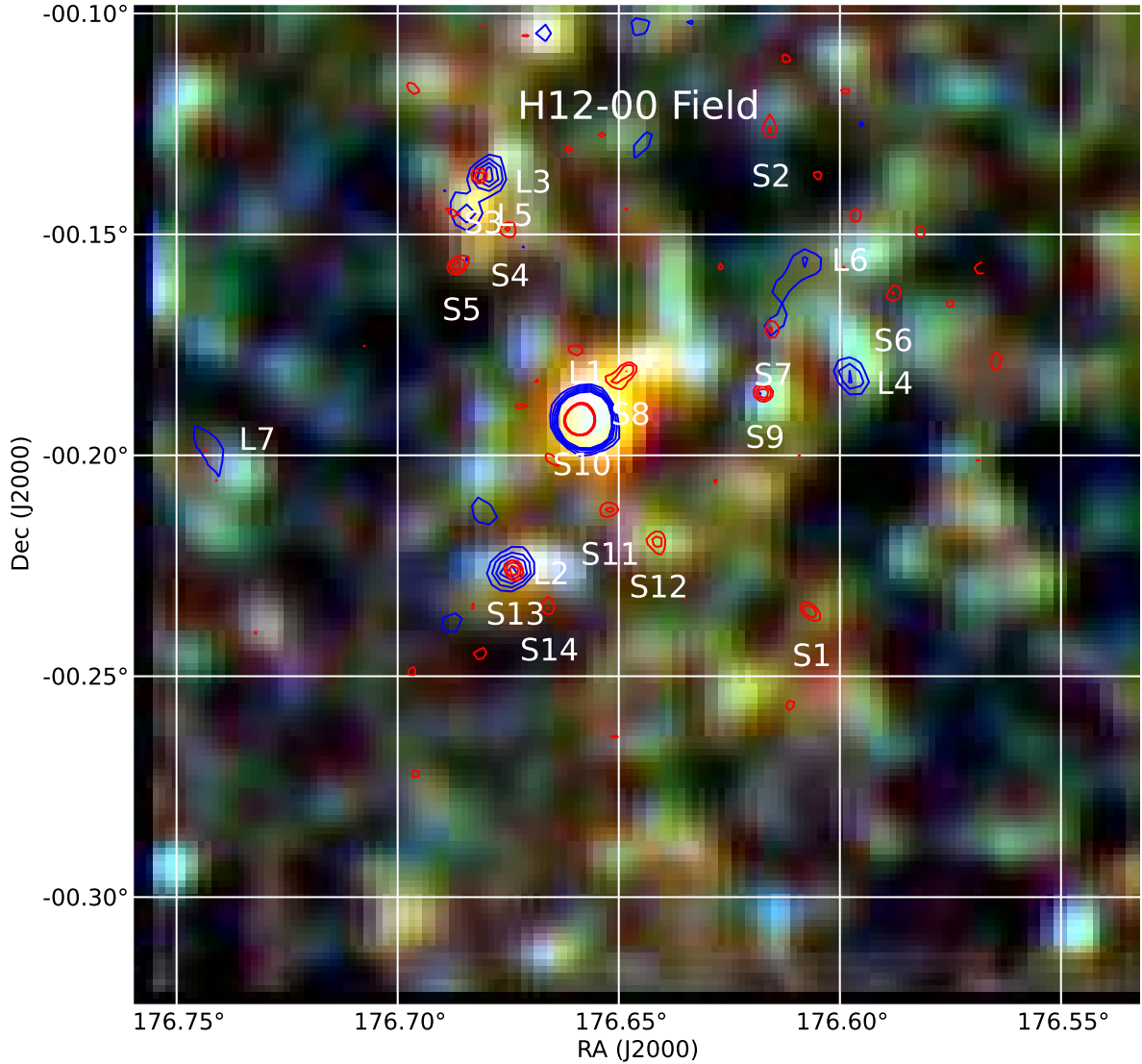, width=16cm}
\caption{Three colour SPIRE image of the 400 $\times$400 arcsec region around the lensed $z=3.26$ source HATLAS12-00. Blue = 250$\mu$m, Green=350$\mu$m and Red=500$\mu$m, overlaid by contours, at levels of 3, 4, 5, 6 and 7 $\sigma$ from the submm observations, with LABOCA in blue and SCUBA2 in red. Source names added as in Fig. 1. Note that those sources brightest at 870 or 850$\mu$m $\mu$m are associated with the reddest SPIRE sources.}
\end{figure*}

\begin{figure*}
\epsfig{file=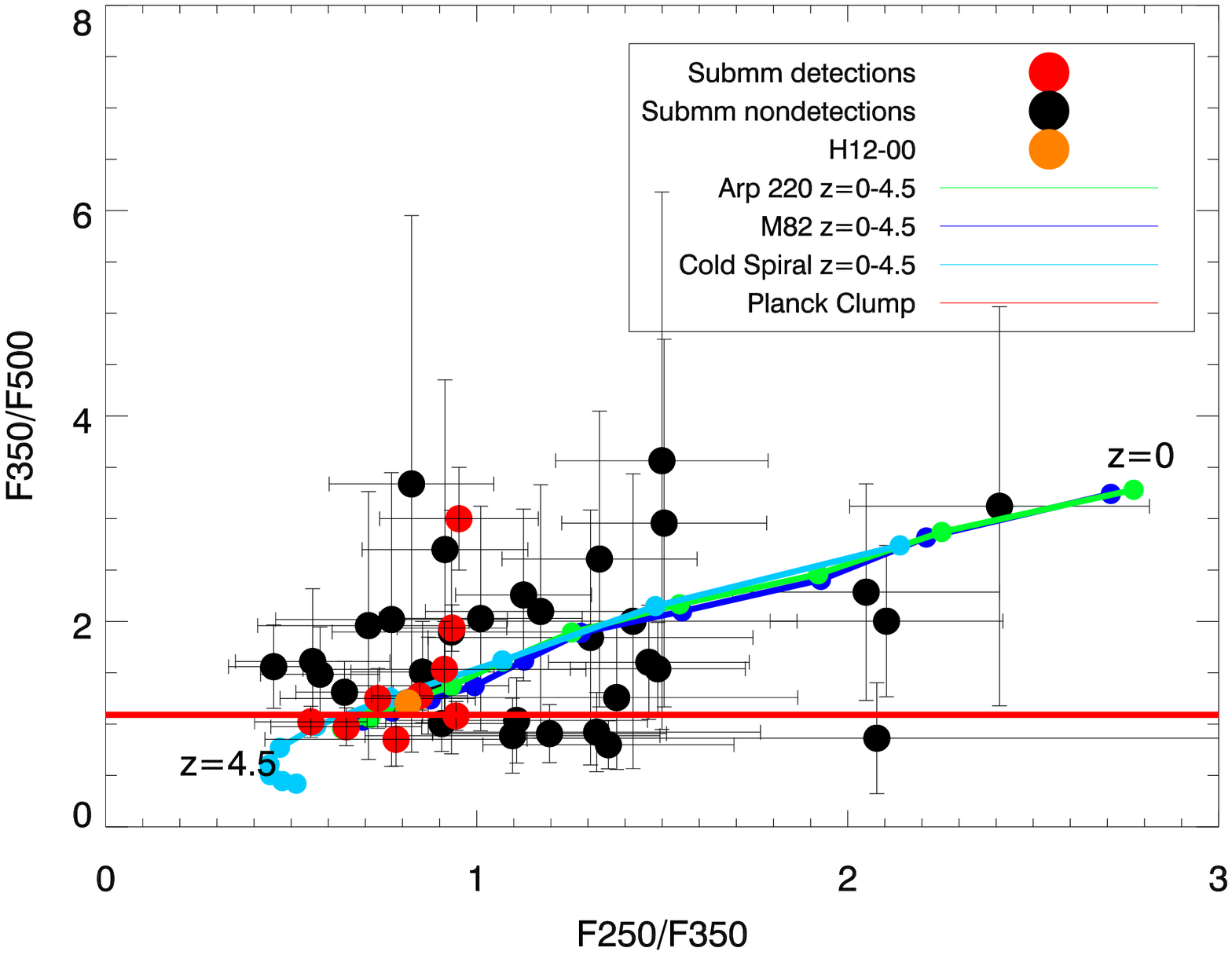, width=12cm, angle=90}
\caption{250$\mu$m/350$\mu$m vs. 350$\mu$m/500$\mu$m flux ratios of SPIRE sources within 5 arcminutes of the lensed source HATLAS12-00 compared to those expected for star forming galaxies as a function of redshift. Submm detected sources are indicated with a red colour, with HATLAS12-00 shown in orange. Black symbols denote SPIRE sources that do not have a $>4\sigma$ submm detection but which have S/N$>$2 in all SPIRE bands. The submm-detected sources are redder than the bulk of the SPIRE sources, and have colours comparable to those of the lensed $z=3.26$ source HATLAS12-00. Some SPIRE-only sources have similarly red colours. The {\em Planck} 857GHz (ie. 350$\mu$m) to 545GHz (ie. 550$\mu$m) colour is shown as a horizontal red line, giving the colour of the clump as a whole. The {\em Planck} colour is comparable to that of the lensed source. Colour tracks are also shown as lines for three template SEDs, two (M82 and Arp220) appropriate for starbursting galaxies, one (cold spiral) appropriate for lower temperature lower luminosity systems. These cover the redshift range from $z=0$ to $z=4.5$. Dots along these lines occur at intervals of 0.5 in redshift.}
\end{figure*}

The {\em Herschel} and {\em Herschel}-submm colours of our submm detected sources are all quite similar to the colours of the $z=3.26$ lensed source. There is thus a temptation to conclude that the submm companions of HATLAS 12-00 all lie at a similar redshift. However, there is also the possibility that the clump of sources, while at high redshift, is not physically associated with the lensed source, as well as the possibility that some, or maybe all, of the companions are cooler objects at a significantly lower redshift. If we take the 20K dust temperature of the cold spiral SED as a likely lower limit to the dust temperature of a galaxy - something that appears to be the case for the vast majority of local galaxies but with a small number of exceptions (Rowan-Robinson \& Clements, 2015) - then the likely lower redshift limit for these sources is z$>$1 and for most z$>1.5$. These limits largely come from the {\em Herschel}-submm colours.

It is also worth noting that eight of the nine good companions to HATLAS12-00 have F250/F350 $<$ 1 ie. their SEDs peak in the 350$\mu$m band of SPIRE. The remaining source has an an SED rising to 500$\mu$m. Tens of `350-peaker' sources now have spectroscopic redshifts as a result of followup observations aimed at candidate lenses and other samples (eg. Wardlow et al., 2013; Harris et al., 2012; Bussmann et al., 2013; Casey et al., 2012a, 2012b). The vast majority of those with secure colours and reliable redshifts have redshifts $>>$2. Thus, despite the redshift-temperature degeneracy, an SED peaking at 350$\mu$m seems observationally to be a very good indication of high redshift.

\begin{figure*}
\epsfig{file=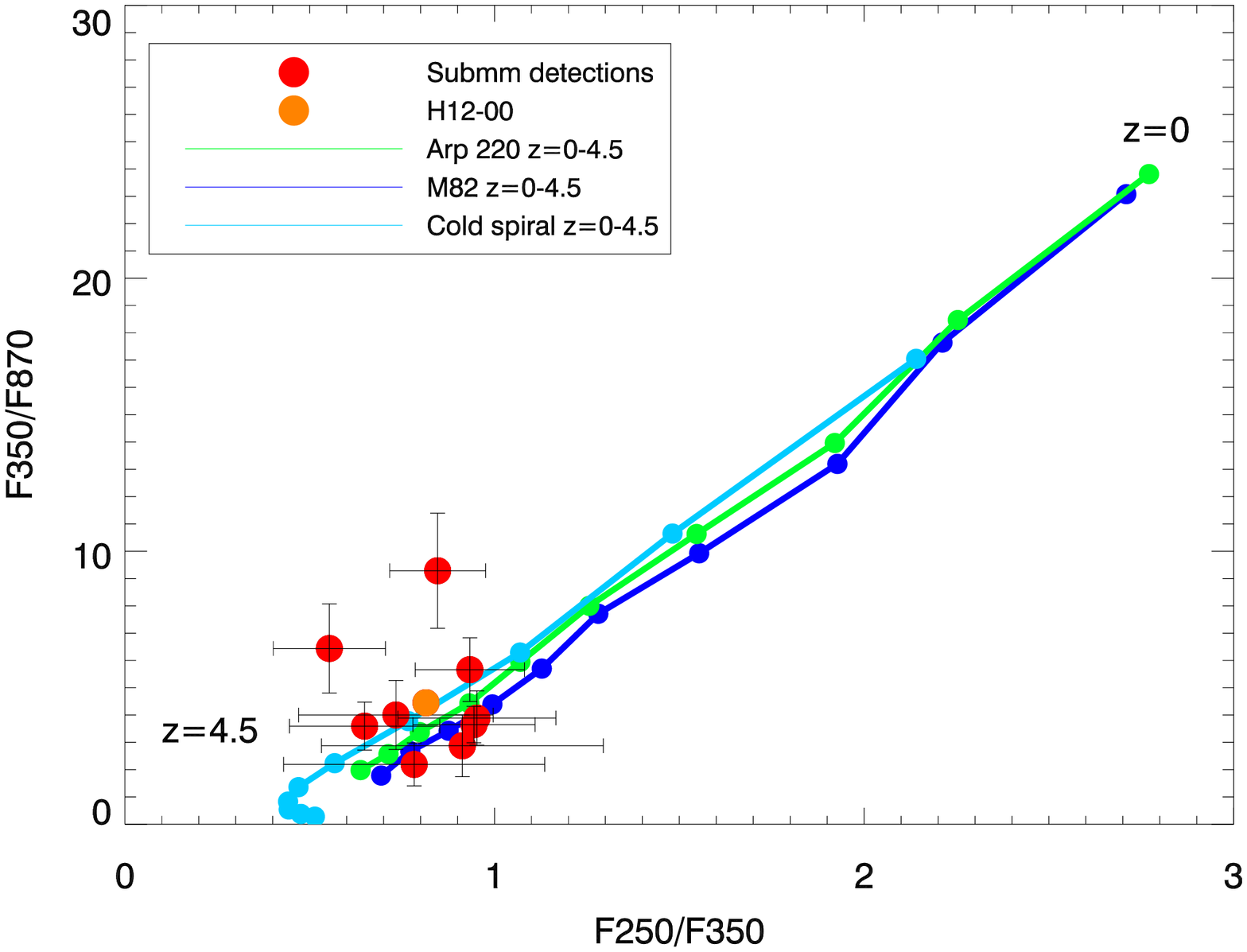, width=12cm, angle=90}
\caption{250$\mu$m/350$\mu$m vs. 350$\mu$m/870$\mu$m flux ratios of the sources detected in the submm classified as good, compared to colour tracks as a function of redshift for star forming galaxies. All these sources have colours similar to the lensed $z=3.26$ source HATLAS12-00, which is the source in this figure with the smallest error bars, consistent with them all lying at a similar redshift. Template colour tracks are also shown, similar to those in Fig. 3.}
\end{figure*}

\subsection{Multiwavelength Cross Identifications}

SMA observations of sources LABOCA\_2, 3, 4 \& 5 have found a source corresponding to LABOCA\_2, with a flux of 4.2$\pm$1mJy (a 4.2 $\sigma$ detection). The location of this SMA source is well matched to the LABOCA and SCUBA2 positions (see Fig. 5).  The non-detections for LABOCA\_3, LABOCA\_4 and 5 are compatible with their expected fluxes based on their colours and SED template fitting - we would at best only have expected to detect them at $\sim$ 3$\sigma$ given the noise achieved at the SMA, and indeed some marginal detections at this level are found (see Section 3.4).

The position for the SMA counterpart to LABOCA\_2 is coincident with a source in our {\em Spitzer} 3.6 and 4.5 $\mu$m images of this field (see Fig. 5 Left). The properties of this cross identification are summarised in Table 5. Examination of SDSS images at the position of this source fails to find any optical counterpart, implying that the $i$-band magnitude of the source is fainter than the 21.3 magnitude $i$-band limit of SDSS. Deeper optical observations of this source obtained with white light (ie. filterless) observations at the WHT also fail to find an optical counterpart. The 5 $\sigma$ detection limit for these observations is estimated to be $i_{AB} < 24.6$ (Fu et al., 2012). This source is detected in our ISAAC J and K' observations of a $\sim$150x150 arcsec field around the position of LABOCA\_2. J and K' magnitudes for this source were extracted with apertures matching those of the IRAC photometry, and found to be J$_{vega}$ = 22.6$\pm$0.15 and K$_{vega}$ = 20.3$\pm$0.15. The colours of this source would classify it as a Distant Red Galaxy (DRG; Franx et al., 2003), a class of object already known to often be the near-IR counterparts of high redshift ($z>2$) submm sources (Dannerbauer et al., 2004).

In Fig. 6 we plot the observed optical/near-IR SED of LABOCA\_2, including the IRAC and ISAAC detections and the WHT upper limit. We perform a photometric redshift analysis of these fluxes using the template SEDs derived by Berta et al. (2013) for {\em Herschel} sources and the wider range of optical/near-IR templates provided by Bruzual \& Charlot (2003). A variety of templates can fit the data and provide a range of acceptable redshifts from 2.8 to 3.9 for the Berta templates and at $z>2$ for the Bruzual \& Charlot models. None of the templates, however, provides an acceptable match to the observed SED at redshifts $\sim$1. We thus conclude that the companion source LABOCA\_2, at least, is much more likely to lie at a redshift comparable to that of the lensed source than that of the lower redshift lens. In Fig. 6 we illustrate this by comparing the best fit Berta template SEDs for $z=3.26$, matching that of the lensed source and representing a high redshift solution, with that for $z=1.22$, matching the redshift of the foreground lensing sources and representing a low redshift solution. We also show the $\chi^2$ values for these fits. The high redshift SED is clearly the better fit, with a reduced $\chi^2$ of 3.9 compared to 23.6 for the low redshift solution, though the J band flux is somewhat higher than the template would suggest. Combining the optical/near-IR data with the far-IR/submm fluxes and fitting with the same Berta et al. (2013) templates provides a similar result. 

Examination of the ISAAC images shows that LABOCA\_2 appears to be extended on scales of a few arcseconds (see Fig. 5 Right). The source seems to have a bimodal flux distribution, suggestive of a merger-like morphology though other interpretations are possible. However, deeper observations are required to confirm this.

\begin{figure*}
%\begin{tabular}
\epsfig{file=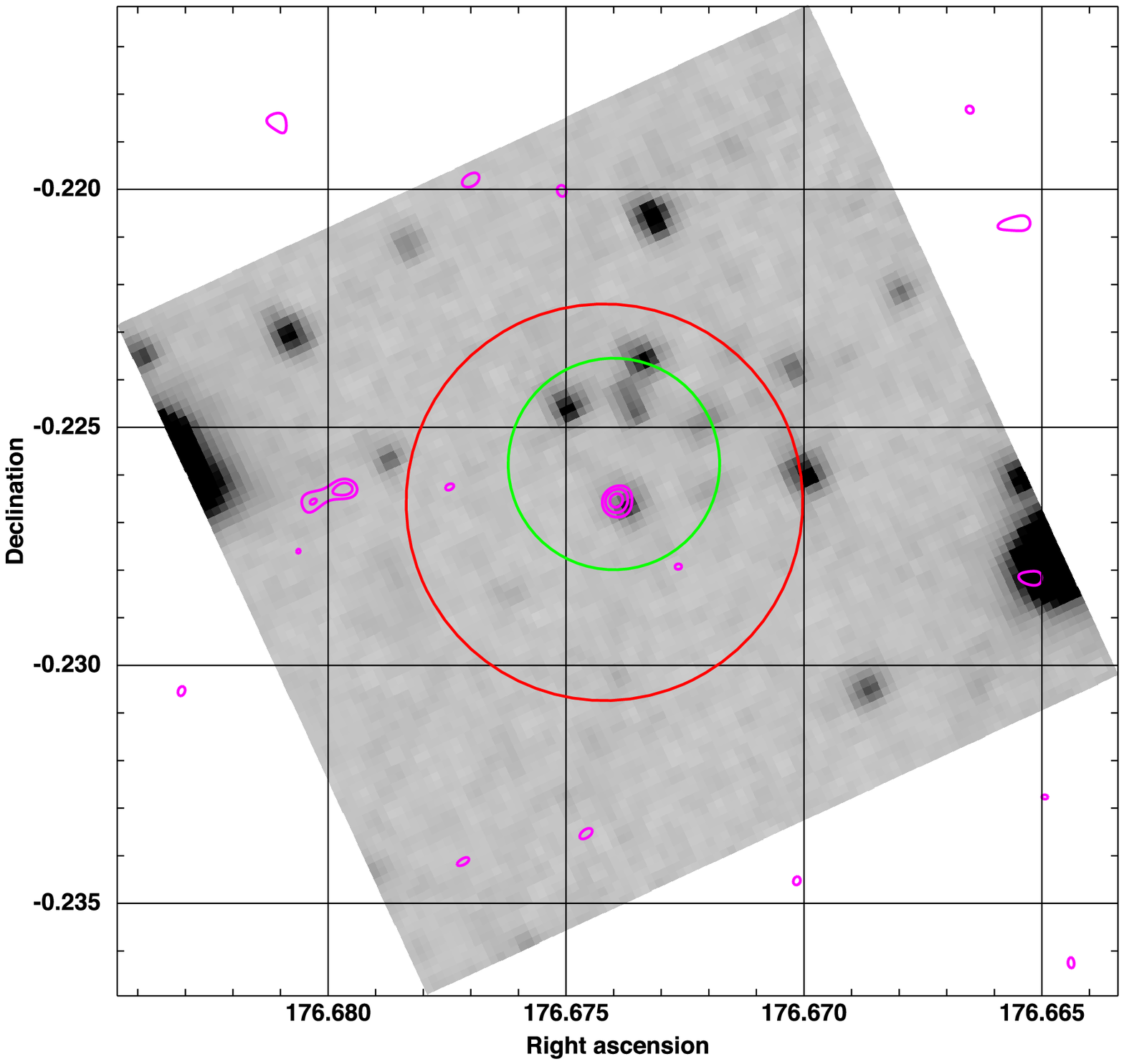,width=11.5cm,angle=0}\\
\epsfig{file=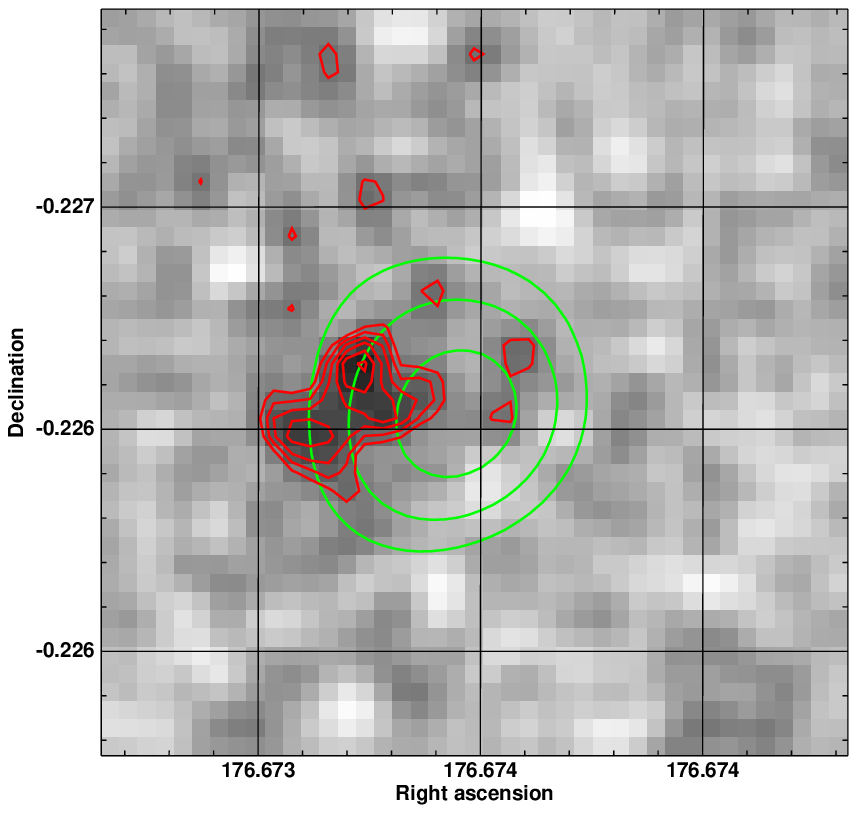, width=11.5cm, angle=-0}
%\end{tabular}
\caption{{\bf Top:} Identification of LABOCA\_2 with an IRAC source. The greyscale image is a segment of the larger IRAC 3.6$\mu$m image. The magenta contours are from the SMA, at levels of 2, 3 and 4 $\sigma$, the red circle is the LABOCA beam at the position of the source, and the green circle is the SCUBA2 beam at the position of the matching source. The correct IRAC identification for this source is clear once the SMA contours are considered. {\bf Bottom:} SMA contours for the LABOCA\_2 source (green) overlaid on a K' band image of the source obtained from ISAAC (greyscale and red contours). The image is 6 x 6 arcseconds in size with 0.15 arcsecond pixels. The original image has been smoothed with a gaussian of width two pixels to enhance the S/N. Note that the K' band counterpart to the SMA source is extended and seems to have a bimodal morphology, suggestive of a galaxy interaction or merger. As can be seen the z=1.22 SED is a poorer fit to the data. For both images coordinates are given in degrees J2000.}
\end{figure*}

\begin{figure}
\epsfig{file=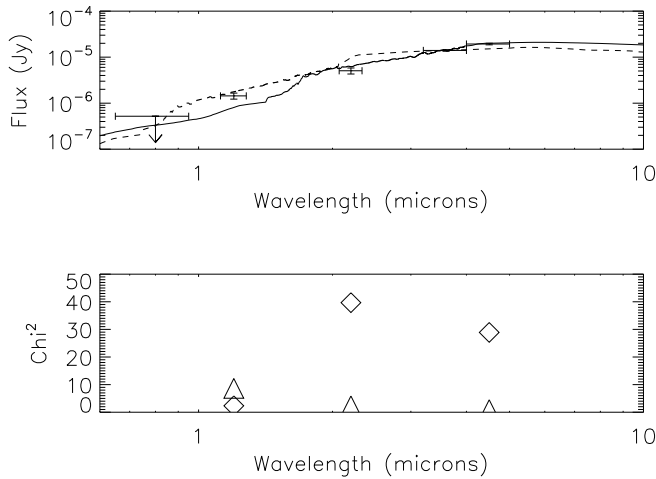,width=7cm, angle=90}
\caption{{\bf Top: }Observed fluxes of LABOCA\_2 from IRAC, ISAAC and the WHT compared to the best fitting Berta et al (2013) template SEDs at z=3.26 (solid) and z=1.22 (dashed), respectively matching the redshift of the lensed source and putative cluster, and the foreground galaxies that are lensing HATLAS12-00. The templates are normalised to the observed flux at 3.6$\mu$m from IRAC. {\bf Bottom:} The $\chi^2$ derived from comparing these fits to the data for the z=1.22 model (diamonds) and the z=3.26 model (triangles).}
\end{figure}

\begin{table*}
\begin{tabular}{ccccccccc}\hline
Source & RA&Dec&F$_{1100}$&Optical&J&K'&[3.6]&[4.5]\\ 
&Deg J2000&Deg J2000&mJy&AB&Vega&Vega&Vega&Vega\\ \hline
LABOCA\_2&176.6736&-0.2266&4.2$\pm$1.1&$i_{AB}>24.6$&22.6$\pm$0.18&20.3$\pm$0.15&18.26$\pm$0.04&17.42$\pm$0.04\\ \hline
%LABOCA\_3&176.680&-0.13736&3.9$\pm$1.3&19.79$\pm$0.14&19.27$\pm$0.17&0.52$\pm$0.2\\ \hline
\end{tabular}
\caption{Data for the source detected by the SMA and cross identified with a {\em Spitzer} IRAC source. The magnitudes of the source in the IRAC bands are isophotal corrected total magnitudes and given in the Vega system.}
\end{table*}

\section{Discussion}

\subsection{The Number of Sources in the Overdensity}

We have found a significant overdensity of both SPIRE and SCUBA2 sources in the region around the z=3.26 gravitationally lensed far-IR luminous galaxy HATLAS12-00. While the overdensity is highly significant, the underlying field source counts are sufficiently high that it is unlikely that all of the sources around HATLAS12-00 are members of the overdensity. The probability that a given number of objects are part of the overdensity, as opposed to being from the field population, can be calculated from the field source counts using Poisson statistics, under the assumption that field sources are distributed at random. The probability that there are $n$ overdensity sources out of $M$ detected sources is thus:
\begin{equation}
p(n|M,\mu) = \frac{ \left [\frac{\mu^{(M-n)}}{(M-n)!}\right] e^{-\mu}}{\sum_{i=0}^{i=n} { \left[ \frac{\mu^{(M-i)}}{(M-i)!}\right]}e^{-\mu}}
\end{equation}
where $\mu$ is the number of field sources that would be predicted in the region of consideration on the basis of the source counts.

To illustrate this for the current case we use the CLS field counts from SCUBA2, and restrict ourselves to examining the nine good companion sources which lie within a 220 arcsecond distance from HATLAS12-00. The field counts would predict 1.7 sources in such a region. The resulting probabilities for the number of overdensity sources is shown as a histogram in Fig. 7. As can be seen, the predicted number of overdensity sources is fairly uncertain, with the most likely numbers ranging from 6 to 9. 
%A similar analysis for all submm detected sources within 301 arcseconds of HATLAS12-00 (see section 4.3.2) finds that the most likely number of overdensity sources is 11 with 9 to 13 having probabilities $>$0.1.

\begin{figure}
\epsfig{file=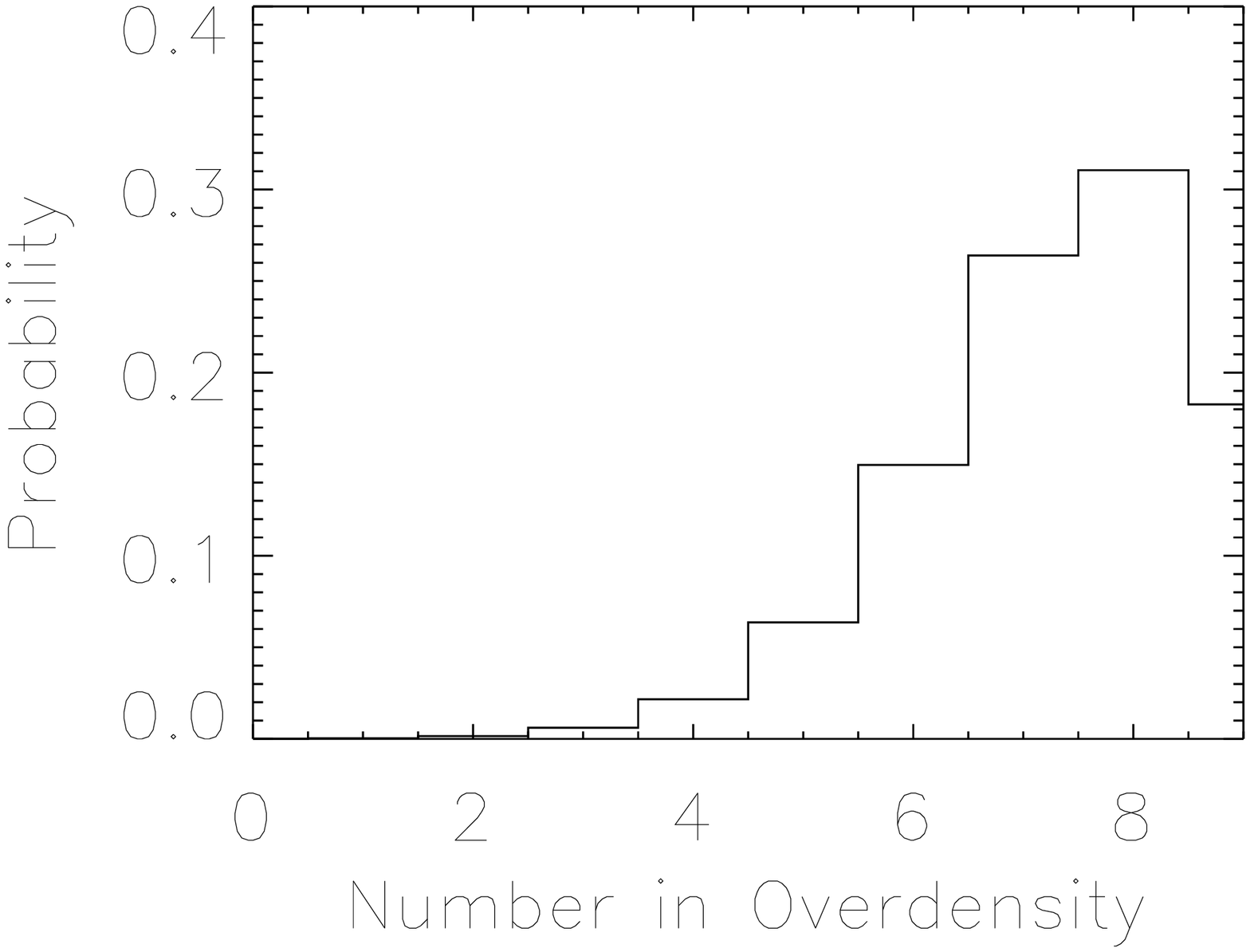,width=7cm, angle=90}
\caption{The number of sources associated with the overdensity rather than the field vs. the probability of this occurring, calculated using Equation 1 and the numbers appropriate for good sources within 220 arcseconds of HATLAS12-00.}
\end{figure}

\subsection{The Nature of the Overdensity}

The foregoing analysis suggests that a significant fraction of the sources we have detected around HATLAS12-00 are likely to lie in an overdensity of objects. It is a natural step from this result to think that these objects are part of a single coherent physical structure, possibly a cluster or protocluster of galaxies. Similar arguments have been applied to the {\em Herschel} and submm source overdensities found by others, including Stephens et al. (2010), Noble et al. (2013), Valtchanov et al. (2013), and Dannerbauer et al. (2014). The similarity of the SPIRE and submm colours of these sources lends credence to this suggestion, but we must be careful since there is a long history of apparent overdensities being identified with single structures when they are in fact chance superpositions of two or more weaker overdensities that just happen to lie along the line of sight (eg. Sutherland \& Efstathiou, 1991). We can assess this possibility by looking at the expected number of clusters that would lie in the volume observed in this study. The cluster X-ray luminosity function down to $\sim 1 \times 10^{43}$ergs s$^{-1}$cm$^{-2}$, appropriate for poor clusters or galaxy groups (Koens et al. (2013)), is known out to $z\sim1$ with reasonable accuracy. Since this is expected to decline to higher redshifts it can be used to calculate conservative upper limits to the expected number of clusters found in any $z>1$ survey. If we were to assume that the HATLAS 12-00 field was selected at random, then the total volume along the line of sight that we have observed out to $z=3.26$ is 3.8$\times 10^5$Mpc$^3$ and we would predict an average of $\sim$ 0.38 poor clusters or groups to lie in this volume. Assuming Poisson statistics, one group/cluster would be found with a probability of 0.26, two with 0.05 and three or more structures with 0.007. There is thus a small ($\sim$6\%) but non-negligible chance that our overdensity might be enhanced by a superposition effect.

Obtaining spectroscopic redshifts for the {\em Herschel} and submm sources we have uncovered is necessary for properly understanding this overdensity. Unfortunately such data is not yet available, but we can get some hints as to what might be going on by applying far-IR photometric redshift techniques, similar to those used by Noble et al. (2013) in studying a submm source overdensity at a suggested redshift of $z\sim3$ which they uncovered behind the $z=0.9$ supercluster RCS 231953+00. This process produces a similar result to that found by comparing the colour tracks of template sources to the observed colours of our sources: the higher luminosity, higher dust temperature templates imply redshifts $z\sim$3, with implied luminosities of $7 - 14 \times 10^{12} L_{\odot}$ for the sources we classify as good, while the lower temperature, lower luminosity templates imply redshifts closer to $z\sim1$ with luminosities $1 - 2 \times 10^{12} L_{\odot}$. Noble et al. (2013) argued that lower temperature, lower redshift solutions were not compatible with the empirical temperature-luminosity relation derived by Roseboom et al. (2012) for {\em Herschel} selected sources. However, further work since then (eg. Casey et al. 2012b) has shown that there is a significant amount of scatter in this relation. This weakens any conclusion that can be drawn from such an analysis. While higher redshifts for our sources are still somewhat favoured by the T-L relation in Casey et al. (2012b) the possibility that some of our sources are cool outliers in this relation cannot be discounted. Our analysis of the optical/near-IR SED of the one source where we have a clear identification, LABOCA\_2 (see Section 4.5), seems to favour higher redshifts, but without enough precision to differentiate between $z\sim2$ and $z\sim3$.

\section{Conclusions}

The lensed $z_{spec}=3.26$ dusty starburst galaxy HATLAS12-00 was found to be associated with a {\em Planck} ERCSC source by Herranz et al. (2013), leading to the suggestion that it is part of a galaxy cluster or protocluster, several of whose members are undergoing a starburst. The presence of a number of {\em Herschel} sources near to HATLAS12-00 with similarly red SPIRE colours to the lens supports this idea. We obtained submm observations of an $\sim$11 arcmin diameter field around HATLAS12-00 using both the LABOCA and SCUBA2 submm mappers. As well as detecting the lensed source itself, we detect 17 other submm sources in this field at 4$\sigma$ significance or greater, with fluxes from 6.4 to 15mJy at 850 and/or 870$\mu$m. The number of such submm sources found in this field represents a significant overdensity, one that would occur at random in the field with a probability ranging from 7.6$\times 10^{-3}$ to 3.8$\times 10^{-4}$, depending on the reliability of the sources included in the calculation. There is also an overdensity of {\em Herschel} sources around HATLAS12-00 which has a probability of 5$\times 10^{-4}$ of arising at random.
These low probabilities argue for a physical association between these sources, suggesting that they are lie in a cluster, a proto-cluster or some other large scale structure at similar redshifts. Whether this structure is related to the $z=3.24$ lensed source is unclear. The {\em Herschel} and submm colours of the sources are red, suggesting that the sources are either warm high luminosity systems at $z>2$ or cooler systems at a redshift closer to 1. Multiwavelength followup observations identify one of the companions to HATLAS12-00 with a source detected by {\em Spitzer} at 3.6 and 4.5 $\mu$m, and in subsequent near-IR imaging using the VLT. Photometric redshift analysis of the optical/NIR SED of this source indicates a redshift $>2$, potentially similar to that of HATLAS12-00 but not compatible with redshifts $\sim$1. Further followup observations including optical/near-IR imaging and spectroscopy are needed to firmly establish the nature of the overdensity we have found and to examine the tantalising possibility that this might be a galaxy cluster/protocluster at z$\sim$3.26.

%Eleven of the submm sources are cross identified with {\em Herschel} sources, allowing us to examine their combined SPIRE and submm colours. We find that these sources all have similar colours to the lensed source, and photometric redshifts, albeit with significant uncertainties,  consistent with the idea that they are lying at the same redshift.  Far infrared luminosities are derived for the submm sources, assuming that they all lie at $z=3.26$, and for HATLAS12-00 after correcting for the lensing magnification derived by Fu et al. (2012). We find that the sources would have L$_{FIR} \sim 10^{12-13} L_{\odot}$ if at $z=3.26$, suggesting that this is a cluster or larger structure containing multiple star forming U/HLIRGs. We derive a total star formation rate for this putative cluster of  25000---30000 $M_{\odot}$yr$^{-1}$. 

%We compare the SFRD in this suspected cluster to that of lower redshift clusters in the literature and to the groups of submm sources that Stevens et al. (2010) found were associated with X-ray selected quasars. We find that the cluster SFRD from z$\sim$1.5 to z$\sim$3 is roughly constant, but that it falls rapidly at lower redshifts. The evolution of cluster SFRD is a largely unexamined aspect of cluster galaxy evolution and may be a useful new tool for comparison to models of galaxy and large scale structure formation, especially since the prospects for finding further high redshift star forming clusters through the combination of {\it Planck} and {\it Herschel} data are very good.

{\bf Acknowledgements}
The {\em Herschel}-ATLAS is a project with {\em Herschel}, which is an ESA space observatory with science instruments provided by European-led Principal Investigator consortia and with important participation from NASA. The H-ATLAS website is http://www.h-atlas.org/. Based in part on observations obtained with {\em Planck} (http://www.esa.int/{\em Planck}), an ESA science mission with instruments and contributions directly funded by ESA Member States, NASA, and Canada. This publication is also based in part on data acquired with the Atacama Pathfinder Experiment (APEX). APEX is a collaboration between the Max-{\em Planck}-Institut fur Radioastronomie, ESO, and the Onsala Space Observatory. The James Clerk Maxwell Telescope has historically been operated by the Joint Astronomy Centre on behalf of the Science and Technology Facilities Council of the United Kingdom, the National Research Council of Canada and the Netherlands Organisation for Scientific Research. The Submillimeter Array is a joint project between the Smithsonian Astrophysical Observatory and the Academia Sinica Institute of Astronomy and Astrophysics and is funded by the Smithsonian Institution and the Academia Sinica. This work is based in part on observations made with the Spitzer Space Telescope, which is operated by the Jet Propulsion Laboratory, California Institute of Technology under a contract with NASA. Based in part on observations made with ESO Telescopes at the La Silla Paranal Observatory under programme ID 290.A-5043. Support for this work was provided by NASA through an award issued by JPL/Caltech.The Italian co-authors acknowledge financial support from ASI/INAF agreement 2014-024-R.0. The Spanish co-authors acknowledge financial support by the Ministerio de Ciencia e Innovacion, AYA2012-39475-C02-01, and Consolider-Ingenio 2010, CSD2010-00064. J.G.N. acknowledges financial support from the Spanish MINECO for a ÒRamon y CajalÓ fellowship and from the Spanish CSIC for a JAE-DOC fellowship, co-funded by the European Social Fund. E. Ibar acknowledges funding from CONICYT/FONDECYT postdoctoral project N$^\circ$:3130504. This work was also supported in part by funding from STFC. LD, RJI, IO and SJM acknowledge support from the European Research Council (ERC) in the form of Advanced Investigator Program, COSMICISM. The Dark Cosmology Centre is funded by the Danish National Research Foundation.
DLC would like to thank Douglas Scott and Andrew Jaffe for useful discussions.
\\~\\

\end{document}